\documentclass[letterpaper, 10 pt, conference]{ieeeconf}  
\IEEEoverridecommandlockouts
\pdfoutput=1
\let\chapter\section

\renewcommand\dbltextfloatsep{5pt}
\renewcommand\dblfloatsep{5pt}
\renewcommand\textfloatsep{5pt}
\usepackage{cite, graphicx,url,setspace}
\usepackage{subfig}
\usepackage[format=plain,labelsep=period,font=footnotesize]{caption}

\usepackage[linesnumbered,ruled]{algorithm2e}
\usepackage{amssymb, color}
\usepackage{bm}
\usepackage{amsmath}
\usepackage{CJK}
\newtheorem{assumption}{Assumption}

\newtheorem{remark}{Remark}

\let\labelindent\relax
\usepackage{enumitem}

\usepackage[figuresright]{rotating}

\allowdisplaybreaks
\begin{document}
\bstctlcite{IEEEexample:BSTcontrol}
\title{\LARGE \bf Stochastic Feedback Control of Systems\\ with Unknown Nonlinear Dynamics$ \tiny^{\tiny{*}} $}

\author{Dan Yu$^{1}$, Mohammadhussein Rafieisakhaei$^{2}$ and Suman Chakravorty$^{1}$
\thanks{*This material is based upon work partially supported by NSF under Contract Nos. CNS-1646449 and Science \& Technology Center Grant CCF-0939370, the U.S. Army Research Office under Contract No. W911NF-15-1-0279, and NPRP grant NPRP 8-1531-2-651 from the Qatar National Research Fund, a member of Qatar Foundation, AFOSR contract Dynamic Data Driven Application Systems (DDDAS) contract FA9550-17-1-0068 and NSF NRI project ECCS-1637889.}
\thanks{$^{1}$D. Yu and S. Chakravorty are with the Department of Aerospace Engineering, and $^{2}$M. Rafieisakhaei Electrical and Computer Engineering, Texas A\&M University, College Station, Texas, 77840 USA. \{\tt\small yudan198811@hotmail.com, mrafieis, schakrav@tamu.edu\}}%
}

\maketitle
\begin{abstract}
This paper studies the stochastic optimal control problem for systems with unknown dynamics. First, an open-loop deterministic trajectory optimization problem is solved without knowing the explicit form of the dynamical system. Next, a Linear Quadratic Gaussian (LQG) controller is designed for the nominal trajectory-dependent linearized system, such that under a small noise assumption, the actual states remain close to the optimal trajectory. The trajectory-dependent linearized system is identified using input-output experimental data consisting of the impulse responses of the nominal system. A computational example is given to illustrate the performance of the proposed approach. 
\end{abstract}
\section{Introduction}
Stochastic optimal control problems, also known as Markov Decision Processes (MDPs) have found numerous applications in the Sciences and Engineering. In general, the goal is  to control a stochastic system so as to minimize the expected running cost of the system. 
It is well known that the global optimal solution for MDPs can be found by solving the Hamilton-Jacobi-Bellman (HJB) equations \cite{dp_bertsekas}. The solution techniques can be further divided into model-based and model-free techniques according to whether the method uses an analytical model of the system, or it uses a black box simulation model or real experiments.

In model-based techniques, many methods rely on a discretization of the underlying state and action space \cite{dp_num}, and hence, run into the curse of dimensionality, the fact that the computational complexity grows exponentially with the dimension of the state space of the problem. The most computationally efficient approach among these techniques are trajectory-based methods, first described in \cite{bryson}. These methods linearize the nonlinear system equations about a deterministic nominal trajectory and perform a localized version of policy iteration to iteratively improve the policy. For instance, the Differential Dynamic Programming (DDP) \cite{ddp, sddp} linearizes the dynamics and the cost-to-go function around a given nominal trajectory, and designs a local feedback controller using DP. The iterative LQG (iLQG) \cite{ilqg1, ilqg2}, which is closely related to DDP, considers the first order expansion of the dynamics (in DDP, a second order expansion is considered), and designs the feedback controller using Riccati-like equations, which is shown to be computationally more efficient. In both approaches, the control policy is executed to compute a new nominal trajectory, and the procedure is repeated until convergence. 

Alternatively, the Trajectory-optimized LQG (T-LQG) approach \cite{rafieisakhaei2017TlqgICRA,tlqg,Separation} was recently proposed by us, which shows that under a first order approximation of the dynamics and cost-to-go function, a near optimal solution can be found by first solving a deterministic trajectory optimization problem, followed by a linear time-varying closed-loop controller design problem. This separated approach can also be extended to the model-free situation, which is the subject of this paper: we use a gradient descent algorithm, and a Linear Time-Varying (LTV) system identification technique, in conjunction with a black box simulation model of the process in order to accomplish the ``separated" design.

In the model-free case, the most popular approaches in the community are the Adaptive Dynamic Programming (ADP) \cite{adp,adp_3} and  Reinforcement Learning (RL) paradigms \cite{rl_1, rl_3}. They are essentially the same in spirit and seek to improve the control policy for a given black box system by repeated interactions with the environment while observing the system's responses. The repeated interactions, or learning trials, allow these algorithms to construct a solution to the DP equation, in terms of the cost-to-go function, in an online and recursive fashion. Another variant of the RL techniques is the so-called Q-learning method, where the basic idea is to estimate a real-valued $Q(x, a)$ function of states, $ x $, and actions, $ a $, instead of the cost-to-go function, $V(x)$. For continuous state and control space problems, the cost-to-go functions and the Q-functions are usually represented in a functionally parameterized form; for instance, in the linearly parametrized form $Q(x,a) = \theta^T \phi(x,a)$, where $\theta$ is the unknown parameter vector, and $\phi$ is  a pre-defined basis function. 

Multi-layer neural networks may also be used as nonlinearly parameterized approximators instead of the linear architecture above. The ultimate goal of these techniques is the estimation/learning of the parameters $\theta$ from learning trials/repeated simulations of the underlying system. However, the size of the parameter $\theta$ grows exponentially in the size of the state space of the problem without a compact parametrization of the cost-to-go or Q function in terms of the a priori chosen basis functions for the approximation, and hence, these techniques are typically subject to the curse of dimensionality. Albeit a compact parametrization may exist, a priori, it is usually never known.

In the past several years, techniques based on the DDP/iLQG approach \cite{ddp,sddp, RLHD4,RLHD5}, such as the RL techniques \cite{RLHD1, RLHD2} have shown the potential for RL algorithms to scale to higher dimensional continuous state and control space problems, in particular, high dimensional robotic task planning and learning problems. These methods are a localized version of the policy gradient \cite{baxter, sutton3,marbach} technique that seek to directly estimate the feedback policy via a compact parameterization. For continuous state and control space problems, the method of choice is to wrap an LQR feedback policy around a nominal trajectory and then perform a recursive optimization of the feedback law, along with the underlying trajectory, via repeated simulations/iterations. However, the parametrization can still be very large for partially observed problems ($O(d^2))$ where $d$ is the dimension of the state space) or large motion planning problems such as systems governed by partial differential equations wherein the (discretized) state is very high dimensional (millions of states). Furthermore, there are convergence problems with these techniques that can lead to policy chatter \cite{RLHD1}.

\textit{Fundamentally, rather than solve the derived ``Dynamic Programming" problem as in the majority of the approaches above  that requires the simultaneous optimization of the feedback law and the underlying trajectory, our approach is to directly solve the original stochastic optimization problem in the ``separated open loop/closed loop" fashion wherein:  1) we solve an open loop deterministic optimization problem to obtain an optimal nominal trajectory in a model-free fashion, and then 2) we design a closed loop controller for the linearized time-varying system around the nominal trajectory, again in a model-free fashion.}  This ``divide and conquer" strategy is nonetheless theoretically sound as shown in the companion paper \cite{Separation}. 
The primary contributions of the proposed approach are:

1) Compared to other RL and ADP techniques, implementation using the proposed approach is simple. The stochastic optimal control problem is separated into two sub-problems: deterministic open-loop trajectory optimization problem and a linear time-varying system identification problem, and in each sub-problem, standard approaches can be used. The open-loop optimization problem is solved using gradient descent and input perturbations. The linearized system is identified via the time-varying ERA \cite{tv_era}, using the impulse responses of the nominal system, and an LQG controller is designed for the resulting linearized system. All of the above is accomplished by only considering a sequence of open loop impulse responses of the unknown system.

2) Unlike other ADP and RL techniques, we specify a detailed set of experiments to accomplish the closed loop controller design for the unknown nonlinear system. This series of experiments consists of a sequence of input perturbations to collect the impulse response of the system, first to find an optimized nominal trajectory, and then to recover the LTV system corresponding to the perturbations of the nominal system in order to design the LQG controller corresponding to the LTV system.

3) Albeit not covered in this paper due to a lack of space, in general, for large scale systems with partially observed states, the time-varying ERA constructs a reduced order model (ROM) of the LTV system, and hence, results in a reduced order estimator and controller. For example, the computational complexity of designing the LQG controller using the proposed approach is $O(n_r^3)$, while it is $O(n_x^3)$ for the original system, where $n_r$ is the order of the reduced model,  and $n_r \ll n_x$. Therefore, for large scale systems such as partially observed systems and systems governed by PDEs, the online implementation of the LQG policy using the proposed approach is still computationally tractable. 

The rest of the paper is organized as follows. Section \ref{Section 2} formulates the problem. In Section \ref{Section 3}, we propose a separation-based stochastic optimal control algorithm with discussions of implementation. Last, in Section \ref{Section 4}, we  test the proposed approach using the inverted pendulum problem. 

\section{Problem Formulation}\label{Section 2}

Consider a discrete time nonlinear dynamical system:
\begin{align} \label{original system}
x_{k + 1} &= f(x_k, u_k, w_k), \nonumber \\
y_k &= h(x_k ,v_k),
\end{align}
where $x_k \in \mathbb{R}^{n_x}$,  $y_k \in  \mathbb{R}^{n_y}$, $u_k \in \mathbb{R}^{n_u}$ are the state, measurement, and the control vectors at time $k$, respectively, process model $f(\cdot)$ and measurement model $h(\cdot)$ are nonlinear, the process noise $w_k$ and measurement noise $v_k$ are assumed to be zero-mean, uncorrelated Gaussian white noise with covariances $W$ and $V$, respectively. 

\begin{assumption}
We assume the system is fully observed: 
\begin{align}
y_k = h(x_k, v_k) = x_k .
\end{align} 
\end{assumption}

\begin{remark}
We make Assumption 1 to simplify the treatment of the problem; since otherwise, the stochastic control problem needs to be treated as a partially observed MDP (POMDP). This generalization may be done in a reasonably straightforward fashion (later discussed in Remark \ref{Remark: 7}).
\end{remark}

\textbf{Stochastic Control Problem:}
For the system with unknown nonlinear dynamics, $f(\cdot)$ , the optimal control problem is to find the control policies $\pi = \{\pi_0, \pi_1, \cdots, \pi_{N -1} \}$ in a finite time horizon $[0, N]$, where $\pi_k$ is the control policy at time $k$, i.e., $u_k = \pi_k(x_k)$,  to minimize the cost function
\begin{align}\label{cost_sto_orig}
J_s = E(\sum_{k = 0}^{N - 1} (x_k^T Q_k x_k + u_k^T R_k u_k) + x_N^T Q_N x_N),
\end{align}
where $Q_k, Q_N\succ 0$ and $R_k\succeq 0$.

\section{Stochastic Feedback Control Algorithm}\label{Section 3}
We compute a locally optimal solution to the stochastic control problem in a separated open loop/closed loop (SOC) fashion, i.e., we first solve a noiseless open-loop optimization problem to find a nominal optimal trajectory. Next,  we design a linearized closed-loop controller around the nominal trajectory, such that, with existence of stochastic perturbations, the state stays close to the optimal open-loop trajectory. The separation-based approach has always been used by Control Engineers in Aerospace Guidance and Robotics problems  in a heuristic fashion \cite{bryson}. However, our recent companion work \cite{Separation}, using the theory of Large Deviations, shows that this separation, results in a near-optimal policy in the small noise case. Moreover, experimental results confirm its validity for moderate noise levels. 

The open loop optimization problem could be solved using a general nonlinear programming solver without knowing the explicit form of the underlying dynamics, i.e., it only accesses a black box simulation model of the dynamics. Next, we perform small input perturbations about the nominal trajectory in order to obtain the impulse responses of the LTV system governing the perturbations form the nominal trajectory, and identify the resulting linear time-varying derivation system from these impulse responses using time-varying Eigensystem Realization Algorithm (ERA) \cite{tv_era}. We consider quadratic cost functions, and design an LQR controller which results in an optimal linear control policy around the nominal trajectory. We discuss each of the above steps in the following section.

\subsection{{Open Loop Optimization}}\label{sec_open}
Consider the noiseless nonlinear system:
\begin{align}
x_{k  + 1} = f(x_k, u_k, 0),~
y_k = x_k, 
\end{align}
with known initial state $x_0$, and let the $N$-step cost function
\begin{align*}
J_d (x_0, \{u_k\}_{k = 0}^{N -1})\! =\!\!  \sum_{k = 0}^{N - 1}(x_k^T Q_k x_k + u_k^T R_k u_k)  + x_N^T Q_N x_N.
\end{align*}
The open loop optimization problem is to find the control sequence $\{\bar{u}_k \}_{k = 0}^{N - 1}$, such that for a given initial state $x_0$,
\begin{align}
\{\bar{u}_k \}_{k = 0}^{N - 1} = \arg\min_{\{u_k\}_{k =0}^{N - 1}} J_d (&x_0, \{u_k\}_{k = 0}^{N -1}), \nonumber \\
\text{s. t. } x_{k + 1} &= f(x_k, u_k, 0),\nonumber \\
y_k &= x_k.
\end{align}

The  problem is solved using the gradient descent approach \cite{gradient, sim_opt},  and the procedure is illustrated as follows. 

Starting from an initial guess of the control sequence $U^{(0)} \!\!=\!\! \{\!u_k^{(0)} \!\}_{k = 0}^{N \!-\! 1}$, the control policy is updated iteratively via
\begin{align}
U^{(n + 1)} = U^{(n)} - \alpha \nabla_U J_d(x_0, U^{(n)}),
\end{align}
until a convergence criterion is met, where $U^{(n)} = \{u_k^{(n)} \}_{k = 0}^{N - 1}$ denotes the control sequence in the $n^{th}$ iteration, and $\alpha$ is the step size parameter. The gradient vector is defined as:
\begin{align}\label{gradv}
\nabla_U J_d(x_0, U^{(n)}) \!\!=\!\! \begin{pmatrix} \frac{\partial J_d}{\partial u_0}, \frac{\partial J_d}{\partial u_1}, \cdots, \frac{\partial J_d}{\partial u_{N - 1}} \end{pmatrix}\!\!|_{x_0, \{u_k ^{(n)}\}_{k = 0}^{N - 1}},\!\!
\end{align}
and without knowing the explicit form of the cost function, each partial derivative with respect to the $i^{th}$ control variable $u_i$ is calculated as:
\begin{align}\label{pd}
\frac{\partial J_d}{\partial u_i}|_{x_0, U^{(n)}} \!\!=& \frac{1}{h}(J_d(x_0, u_0^{(n)}, u_1^{(n)},\! \cdots\!, u_i^{(n)} \!\!+\!\! h,\! \cdots\!, u_{N - 1}^{(n)}) \nonumber \\
&-\! J_d(x_0, u_0^{(n)}, u_1^{(n)},\! \cdots\!, u_i^{(n)},\! \cdots\!, u_{N - 1}^{(n)})\!),\!\!\!
\end{align}
where $h$ is a small constant perturbation. 

Algorithm \ref{algo_gradient} summarizes the gradient descent approach.

\begin{algorithm}[!t]
\caption{Gradient descent Algorithm\label{algo_gradient}}
\SetKwInOut{Input}{Input}
\SetKwInOut{Output}{Output}
\Input{Design Parameters $U^{(0)} = \{u_k^{(0)} \}_{k = 0}^{N - 1}$, $\alpha, h, \epsilon$}
\Output{Optimal control sequence $\{ \bar{u}_k \}_{k = 0}^{N - 1}$}
$n \gets 0$, $\nabla_U J_d(x_0, U^{(0)}) \gets \epsilon $\;
\While{$\nabla_U J_d(x_0, U^{(n)}) \geq \epsilon $}{
Evaluate the cost function with control $U^{(n)}$\;
Perturb each control variable $u_i^{(n)}$ by $h$, $i = 0, \cdots, N - 1$, and calculate the gradient vector $\nabla_U J_d(x_0, U^{(n)})$ using (\ref{gradv}) and (\ref{pd})\; 
Update the control policy:
$ U^{(n + 1)} \gets U^{(n)} - \alpha \nabla_U J_d(x_0, U^{(n)}) $\;
$n \gets n + 1$\;
}
 $\{\bar{u}_k \}_{k = 0}^{N - 1} \gets  U^{(n)}$.
\end{algorithm}

\begin{remark}
The open loop optimization problem is solved using a black box model of the underlying dynamics, with sequence of input-output tests. Higher order approaches other than gradient descent are possible \cite{sim_opt}, however, for a general system with complex cost functions, the gradient descent approach is easy to implement and is amenable to very large scale parallelization. 
\end{remark}
\subsection{Linear Time-varying System Identification}
We linearize the system (\ref{original system}) around the optimal nominal control and its corresponding state trajectory denoted by $\{\bar{u}_k\}_{k = 0}^{N - 1}$ and $\{\bar{x}_k\}_{k = 0}^N$, respectively, as: 
\begin{align}\label{perturbation system}
\delta x_{k+ 1} &= A_k \delta x_k + B_k \delta u_k + G_k w_k, \nonumber \\
\delta y_k &= C_k \delta x_k + F_k v_k, 
\end{align}
where $\delta x_k = x_k  - \bar{x}_k$ describes the state deviations from the nominal trajectory, $\delta u_k = u_k - \bar{u}_k$ describes the control deviations,  $\delta y_k = y_k - h(\bar{x}_k, 0)$ describes the measurement deviations, and
\begin{align*}
& A_k \!\!=\!\! \frac{\partial f(x, u, w)}{\partial x}|_{\bar{x}_k, \bar{u}_k, 0}, B_k \!\!=\!\! \frac{\partial f(x, u, w)}{\partial u}|_{\bar{x}_k, \bar{u}_k, 0},  \nonumber \\
& G_k \!\!=\!\! \frac{\partial f(x,\! u,\! w)}{\partial w}\!|_{\bar{x}_k, \bar{u}_k, 0},\!
 C_k \!\!=\!\! \frac{\partial h(x,\! v)}{\partial x}\!|_{\bar{x}_k, 0},\!  F_k \!\!=\!\! \frac{\partial h(x,\! v)}{\partial v}\!|_{\bar{x}_k, 0}. 
\end{align*}

Given the LTV system $\{ A_k, B_k, C_k, G_k, F_k \}$  between time $[0, N-1]$,  an LQR controller to track the nominal trajectory could be designed. However, since the dynamics are unknown, we first need to identify the LTV system.

The time-varying ERA is used to construct a state space realization $(\hat{A}_k, \hat{B}_k, \hat{C}_k , \hat{G}_k, \hat{F}_k \}$ of system (\ref{perturbation system}). The state space realization is constructed using input and output experimental data and is shown to be minimal and balanced. The details of the time-varying ERA can be found in \cite{tv_era} and is briefly summarized next. 

Define the generalized Markov parameters $h_{k, j}$ as:
\begin{align}
h_{k, j}= \begin{cases} C_k A_{k - 1} A_{k - 2} \cdots A_{j+1} B_j, & \text{ if } j < k - 1, \\
 C_k B_{k -1},  & \text{ if } j = k - 1, \\
  0, & \text{ if } j > k - 1,
 \end{cases}
\end{align}

and the generalized Hankel matrix as: 
\begin{align}\label{hankel}
\underbrace{H_k^{(p, q)}}_{pn_y \times qn_u} \!\!\!=\!\!\! \begin{pmatrix}\!  h_{k, k-1} \!& h_{k, k-2} \!\!\!& \cdots \!\!\!& h_{k, k - q} 
\!\\\!
h_{k + 1, k - 1} \!\!& h_{k + 1, k -2} \!\!\!& \cdots \!\!\!& h_{k + 1, k-q} \!\\\!
\vdots \!\!& \vdots \!\!\!& \cdots \!\!\!& \vdots \!\\\!
h_{k+ p- 1, k -1} \!\!& h_{k + p - 1, k -2} \!\!\!& \cdots \!\!\!& h_{k + p - 1, k -q} \!\end{pmatrix}\!\!\!,\!\!\!
\end{align}
with design parameters $p$ and $q$. 

The time-varying ERA starts with an estimation of the generalized Markov parameters from input-output data using least squares solution as follows. 

Consider system (\ref{perturbation system}) with zero noise and for simplicity, assume $\delta x_0 = 0$. Run $M$ simulations and in the $i^{th}$ simulation, choose input sequence $\{ \delta u_t^i \}_{t = 0}^k$, and collect the output $\delta y_k^i$. The superscript $(\cdot)^i$ denotes the experiment number.

From the input-output map, the generalized Markov parameters $\{ h_{k, j} \}_{j = 0}^k $ could be recovered via solving the least squares problem: 
\begin{align}\label{gmarkov}
\begin{pmatrix} \delta y_k^1 \!& \delta y_k^2 \!\!\!&\cdots\!\! \!& \delta y_k^M \end{pmatrix} 
 \!\!=&\!\! \begin{pmatrix} 0 \!\!& h_{k, k-1} \!\!& h_{k, k- 2} \!\!\!& \cdots \!\!& h_{k, 0} \end{pmatrix} \nonumber \\
& \times\!\!\! \begin{pmatrix}\!\! \delta u_k^1 \!& 
\delta u_k^2 \!\!\!& \cdots \!\!& \delta u_k^M 
\!\!\\\!\! \delta u_{k -1}^1 \!& \delta u_{k -1}^2 \!\!\!& \cdots \!\!& \delta u_{k - 1}^M 
\!\!\\\!\! \vdots \!\!& \vdots \!\!& \!\!&\vdots \!\!\\ \delta u_{0}^1 \!& \delta u_0^2 \!\!& \cdots \!\!& \delta u_0^M \!\!\end{pmatrix}\!\!,\!\!\!
\end{align} 
where $M$ is a design parameter and is chosen such that the least squares solution is possible. 

After recovering the generalized Markov parameters, two Hankel matrices $H_{k}^{(p, q)}$ and $ H_{k + 1}^{(p, q)}$ are constructed using (\ref{hankel}), and here, the design parameters $p$ and $q$ are chosen such that $\min \{ p n_y, q n_u \} \geq n_x$, and could be tuned for best performance. 

Then we solve the singular value decomposition problem: 
\begin{align}\label{svd}
H_k^{(p, q)} = \underbrace{U_k \Sigma_k^{1/2}}_{O_k^{(p)}} \underbrace{ \Sigma_k^{1/2} V_k^T}_{R_{k-1}^{(q)}}.
\end{align}
Suppose the rank of the Hankel matrix $H_k^{(p, q)}$ is $n_r$, where $n_r \leq n_x$. Then 
$\Sigma_k \in \mathbb{R}^{n_r \times n_r}$ is the collection of all non-zero singular values,  and $U_k \in \mathbb{R}^{p n_y \times n_r}$, $V_k \in \mathbb{R}^{q n_u \times n_r}$ are the corresponding left and right singular vectors. Similarly, $H_{k + 1}^{(p, q)} = O_{k + 1}^{(p)} R_k^{(q)}$.  

Thus, the identified system using time-varying ERA is:
\begin{align}\label{ltv_id}
&\underbrace{\hat{A}_k}_{n_r \times n_r} = (O_{k + 1}^{(p) \downarrow})^{+} O_k^{(p)\uparrow} ,~
\underbrace{\hat{B}_k}_{n_r \times n_u} = R_k^{(q)}(:, 1: n_u), \nonumber \\
&\underbrace{\hat{C}_k}_{n_y \times n_r} = O_k^{(p)}(1: n_y, :), 
\end{align}
where $(\cdot)^{+}$ denotes the pseudo inverse of $(\cdot)$, $O_{k + 1}^{(p) \downarrow}$ contains the first $(p- 1) n_y$ rows of $O_{k + 1}^{(p)}$, and $O_k^{(p)\uparrow}$ contains the last $(p-1) n_y$ rows of $O_k^{(p)}$. Here, we assume that $n_r$ is constant through the time period of interest, which could also be relaxed. 

\begin{remark}
The uncontrollable or unobservable eigenmodes of the dynamical system are not present in the input-output map, and hence, the state space realization using time-varying ERA is balanced in the sense that only the controllable and observable eigenmodes are preserved.  Hence, for systems with high dimensions, such as systems discretized from partial differential equations (PDEs), and with partially observed states, we have $n_r \ll n_x$. Therefore, one major contribution of this work is that we design a reduced order estimator and controller using the identified system, which implies that the computational complexity is reduced significantly. In comparison,  the computational complexity of designing an LQG controller using the identified system is $O((\frac{n_r}{n_x})^3)$ using the full order system. 
\end{remark}

Note that we cannot perturb the system (\ref{perturbation system}) directly. Instead, we identify the generalized Markov parameters as follows. 

Run M parallel simulations with the noise-free system:
\begin{align}\label{noise_free_input}
x_{k + 1}^i &= f(x_k^i, \bar{u}_k + \delta u_k^i, 0), \nonumber \\
y_k^i &= h(x_k^i, 0),
\end{align}
and therefore,
\begin{align}\label{noise_free_output}
\delta y_k^i = y_k^i - h(\bar{x}_k, 0).
\end{align}
where $(\bar{u}_k, \bar{x}_k)$ is the open loop optimal trajectory. Then solve the same least squares problem with (\ref{gmarkov}).

{For simplicity, we assume that the process noise is independent of the state and control variables, and $G_k = I_{n_x \times n_x}$, $F_k = I_{n_y \times n_y}$, while the proposed algorithm is extendable to identify $G_k$ and $F_k$.}

In general, the identified deviation system is: 
\begin{align}\label{rom}
\delta a_{k  +1} &= \hat{A}_k \delta a_k + \hat{B}_k \delta u_k + \hat{G}_k w_k, \nonumber \\
\delta y_k &= \hat{C}_k \delta a_k + \hat{F}_k v_k, 
\end{align}
where $\delta a_k \in \Re^{n_r}$ denotes the reduced order deviation states. 

Algorithm \ref{algo_ltv} summarizes the time-varying ERA. 
\begin{algorithm}[!tb]
\caption{LTV System Identification\label{algo_ltv}}
\SetKwInOut{Input}{Input}
\SetKwInOut{Output}{Output}
     \Input{Design Parameters $M, p, q$}
 \Output{$\{ \hat{A}_k, \hat{B}_k, \hat{C}_k \}$}
 $k \gets 0$\;
 \While{$k \leq N - 1$}{
 Identify generalized Markov parameters with input and output experimental data using
 (\ref{noise_free_input}), (\ref{noise_free_output}), (\ref{gmarkov})\;
 Construct the generalized Hankel matrices $H_{k}^{(p, q)}$, $H_{k + 1}^{(p, q)}$  using (\ref{hankel})\;
 Solve the SVD problem, and construct $\{\hat{A}_k, \hat{B}_k, \hat{C}_k \}$ using (\ref{ltv_id})\;
 $k \gets k + 1$\;
 }
\end{algorithm}

\subsection{Closed-loop Controller Design}
Given the identified deviation system (\ref{rom}), we design the closed-loop controller to  track the optimal nominal trajectory, which is to minimize the cost function
\begin{align}
J_f = \sum_{k = 0}^{N - 1} (\delta \hat{a}_k^T Q_k \delta \hat{a}_k+ \delta u_k^T R_k \delta u_k ) + \delta \hat{a}_N^T Q_N \delta \hat{a}_N,
\end{align}
where $\delta \hat{a}_k$ denotes the estimates of the deviation state $\delta a_k$. For the linear system (\ref{rom}), the separation principle of control theory is used to separate the design of an estimator and a fully observed controller. 

The feedback controller is:
\begin{align*}
\delta u_k = -L_k \delta \hat{a}_k,
\end{align*}
where $\delta \hat{ a}_k$ denotes the estimates from a Kalman observer, and the feedback gain $L_k$ is computed by solving two decoupled Riccati equations: 
\begin{align}\label{r_1}
L_k = (\hat{B}_k^T S_{k + 1} \hat{B}_k + R_k)^{-1} \hat{B}_k^T S_{k + 1} \hat{A}_k, 
\end{align}
where $S_k$ is determined by running the following Riccati equation backward in time: 
\begin{align}\label{r_2}
S_k \!\!=& \hat{A}_k^T S_{k + 1} \hat{A}_k  + Q_k \nonumber \\
& - \hat{A}_k S_{k + 1} \hat{B}_k(\hat{B}_k^T S_{k + 1} \hat{B}_k + R_k)^{-1} \hat{B}_k^T S_{k + 1} \hat{A}_k, 
\end{align}
with terminal condition $S_N = Q_N$. 

The Kalman filter observer is designed as follows: 
\begin{align}\label{k_1}
\delta \hat{a}_{k + 1} =& \hat{A}_k \delta \hat{a}_k + \hat{B}_k \delta u_k \nonumber \\
& + K_{k + 1}(\delta y_{k + 1} - \hat{C}_{k + 1} (\hat{A}_k \delta \hat{a}_k + \hat{B}_k \delta u_k)),
\end{align}
with $\delta y_k = h(x_k, v_k) - h(\bar{x}_k, 0)$, and the covariance of the estimation is:
\begin{align}\label{k_2}
P_{k + 1} =& \hat{A}_k(P_k - P_k \hat{C}_k^T (\hat{C}_k P_k \hat{C}_k^T + \hat{F}_k V \hat{F}_k^T)^{-1} \hat{C}_k P_k) \hat{A}_k^T \nonumber \\
&+ \hat{G}_k W \hat{G}_k^T,
\end{align}
where the Kalman gain is:
\begin{align}
K_k = P_k \hat{C}_k^T (\hat{C}_k P_k \hat{C}_k^T + \hat{F}_k V \hat{F}_k^T)^{-1}. 
\end{align}

\begin{remark}
Albeit we have only considered the fully observed problem in this paper, any implementation will have noisy measurements, and thus, an observer will be required to implement a feedback controller. In our case, it is simply the LQG controller as outlined above which can be conveniently designed using the identified LTV system .
\end{remark}
\subsection{Stochastic Feedback Control Algorithm}
Algorithm \ref{algo_sc} summarizes the Stochastic Control Algorithm.

\begin{algorithm}[!tb]
\caption{Stochastic Feedback Control\label{algo_sc}}
 Solve the deterministic open-loop optimization problem, to obtain the optimal trajectory $\{\bar{u}_k\}_{k = 0}^{N - 1}, \{\bar{x}_k \}_{k = 0}^{N}$\;
 Identify the LTV system using Algorithm \ref{algo_ltv}\;
 Solve the decoupled Riccati equations (\ref{r_1}, \ref{r_2}) for feedback gain $\{ L_k \}_{k = 0}^{N}$\;
 $k \gets 0, \delta \hat{a}_0 \gets 0$, given $ P_0$\;
 \While{$k \leq N - 1$}{
$ u_k \gets \bar{u}_k - L_k \delta \hat{a}_k $,
$ x_{k + 1} \gets f(x_k, u_k, w_k) $,\\
$ y_{k + 1} \gets h(x_{k + 1}, v_{k + 1}) $\;
Update $\delta \hat{a}_k, P_k$ using (\ref{k_1}, \ref{k_2})\;
$k \gets k + 1$.\;
}
\end{algorithm}

\begin{remark}\textit{Extension to Data-Driven Controller Design.}
As mentioned in data-based LQG \cite{dlqg_skelton, dlqg_2} and data-driven MPC control \cite{dd_1, dd_2}, the linear system ($\hat{A}_k, \hat{B}_k, \hat{C}_k, \hat{E}_k, \hat{F}_k)$ need not be identified to design the LQG controller. Once the generalized Markov parameters are recovered, then together with other input-output data matrices, the controller can also be directly designed. The extension to designing such a ``direct" data-driven controller is not covered in this paper.
\end{remark}
\begin{remark}\textit{Discussion of Implementation Issues.}
\begin{itemize}
\item The proposed approach is implemented in an offine-online fashion. Starting from an given initial state, the open-loop trajectory optimization problem and the LTV system identification problem are solved offline, using the input-output experimental data. Then the LQG policy is implemented online.  
\item An initial estimate of the optimal trajectory is required in the proposed approach, which poses the challenge, and results in a sub-optimal path. 
\item The proposed approach is valid under small noise assumption. In practice, with the presence of noise, non-linearities, and unknown perturbations, the actual state deviates from the nominal trajectory during the execution, and if the actual system deviates too much from the nominal trajectory, the linearization becomes invalid. Therefore, once the deviation is greater than some predefined threshold, a replanning starting from the current location can be performed.
\end{itemize}
\end{remark}

\begin{remark}\textit{Extension to Partially Observed MDP (POMDP).}\label{Remark: 7}
{The proposed approach can also be extended to solve the stochastic optimal control problem for a general nonlinear system with unknown dynamics $f(\cdot)$ and $h(\cdot)$. The only difference is that for the partially observed case, we need to solve the open loop optimization problem in belief space instead of state space. Assume that the belief $b_k = (\mu_k, \Sigma_k)$ is approximately Gaussian, where $\mu_k$ and $\Sigma_k$ represent the mean and covariance of the belief state.  The challenge is that we do not have access to the covariance evolution equation. An effective way is to simulate the covariance evolution using an Ensemble Kalman Filter \cite{evensen}.
Then, the  gradient descent approach can be used for solving the open loop belief space optimization problem. We linearize the system around the optimal control sequence $\{\bar{u}_k\}$, with the associated nominal belief state $(\bar{\mu}_k, \bar{\Sigma}_k)$, and follow the same system identification and closed loop controller design procedure.  The extension to POMDPs will be considered in future work. }
\end{remark}

\section{{Computational Results}}\label{Section 4}
We test the method using the inverted pendulum problem \cite{ipend}. The dynamics of the inverted pendulum mounted on a motor driven cart are given as: 
\begin{align}
&\dot{x}_1 = x_2, \nonumber \\
&\dot{x}_2 =  \frac{u \cos x_1 - (M + m) g \sin x_1 + m l (\cos x_1 \sin x_1) x_2^2}{ml \cos^2 x_1 - (M + m)l}, \nonumber \\
&\dot{x}_3  = x_4, \nonumber \\
&\dot{x}_4 = \frac{u + ml (\sin x_1) x_2^2 - m g \cos x_1 \sin x_1}{M + m - m \cos^2 x_1}
\end{align}
with state variables
\begin{align}
x_1 = \theta,~ x_2 = \dot{\theta},~ x_3 = x,~ x_4 = \dot{x},
\end{align}
where $\theta$ is the tilt angle referenced to the vertically upward direction and  $x$ represents the cart position, $u$ is the $x$-directed external force, the cart mass is $M = 2.4 kg$, the pendulum point mass is $m = 0.23 kg$, the rod length is $l = 0.36 m$, and the standard gravity is $g = 9.81 m/s^2$. 

We use a time step of $0.1s$ to discretize the system in time:
\begin{align}
x_{k + 1} &= f(x_k, u_k) +  w_k, \nonumber \\
y_k &= x_k + v_k,
\end{align}
where $f$ is the unknown dynamics, the process noise and measurement noise $w_k, v_k$ are Gaussian white noise with zero mean and covariances $W = 0.01 I_{4 \times 4}$ and $V = 0.01 I_{4 \times 4}$, respectively. 

\begin{figure}[!tb]
\centering
\includegraphics[width = 0.35 \textwidth]{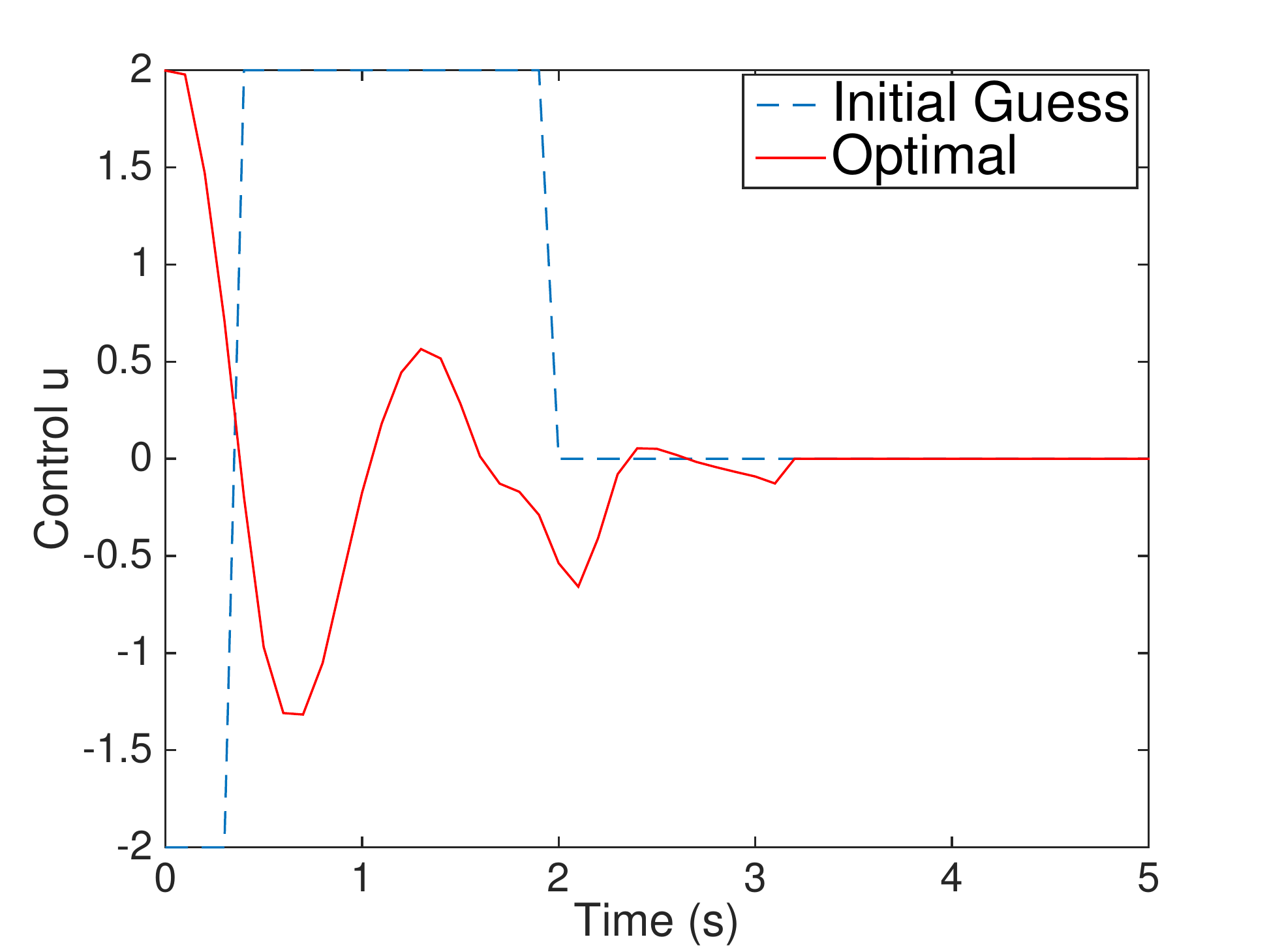}
\caption{Open Loop Optimization Problem.\label{result_open_opt}}\vspace{-5pt}
\end{figure}

\begin{figure*}[htbp]
\centering
   \subfloat{\includegraphics[width= 0.32\textwidth]{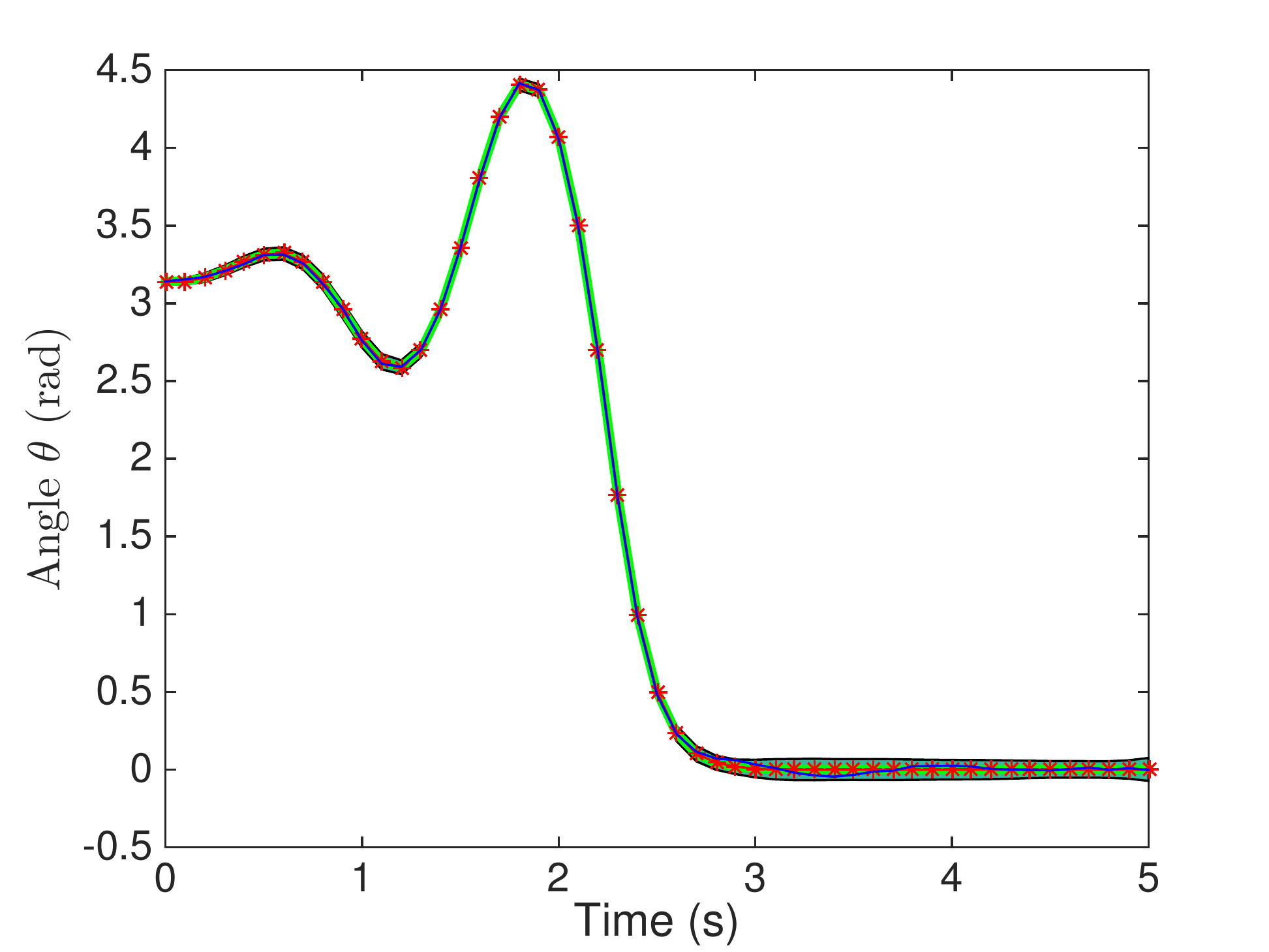}}
   \subfloat{\includegraphics[width= 0.32\textwidth]{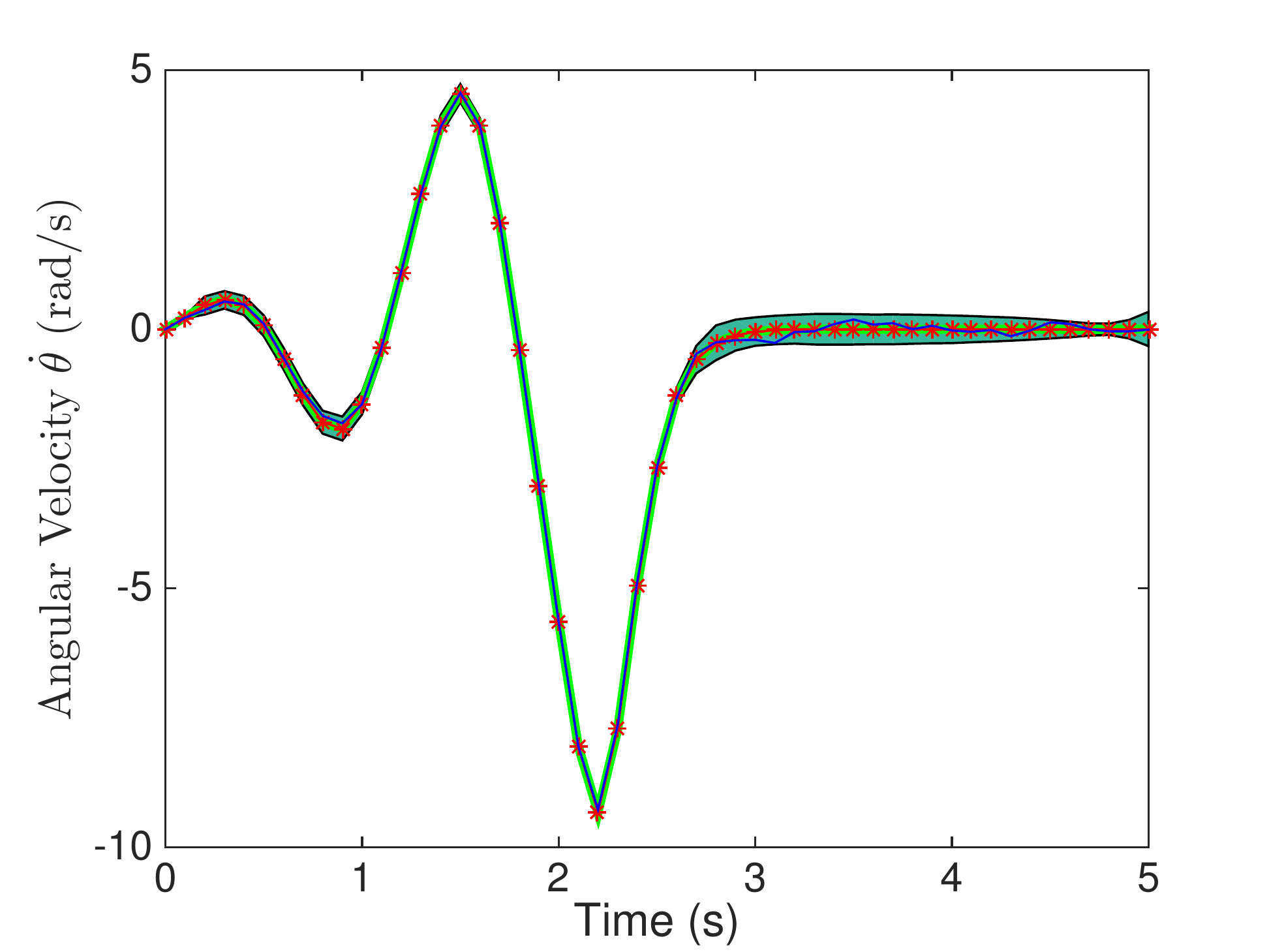}}\\\vspace{-7pt}
   \subfloat{\includegraphics[width= 0.32\textwidth]{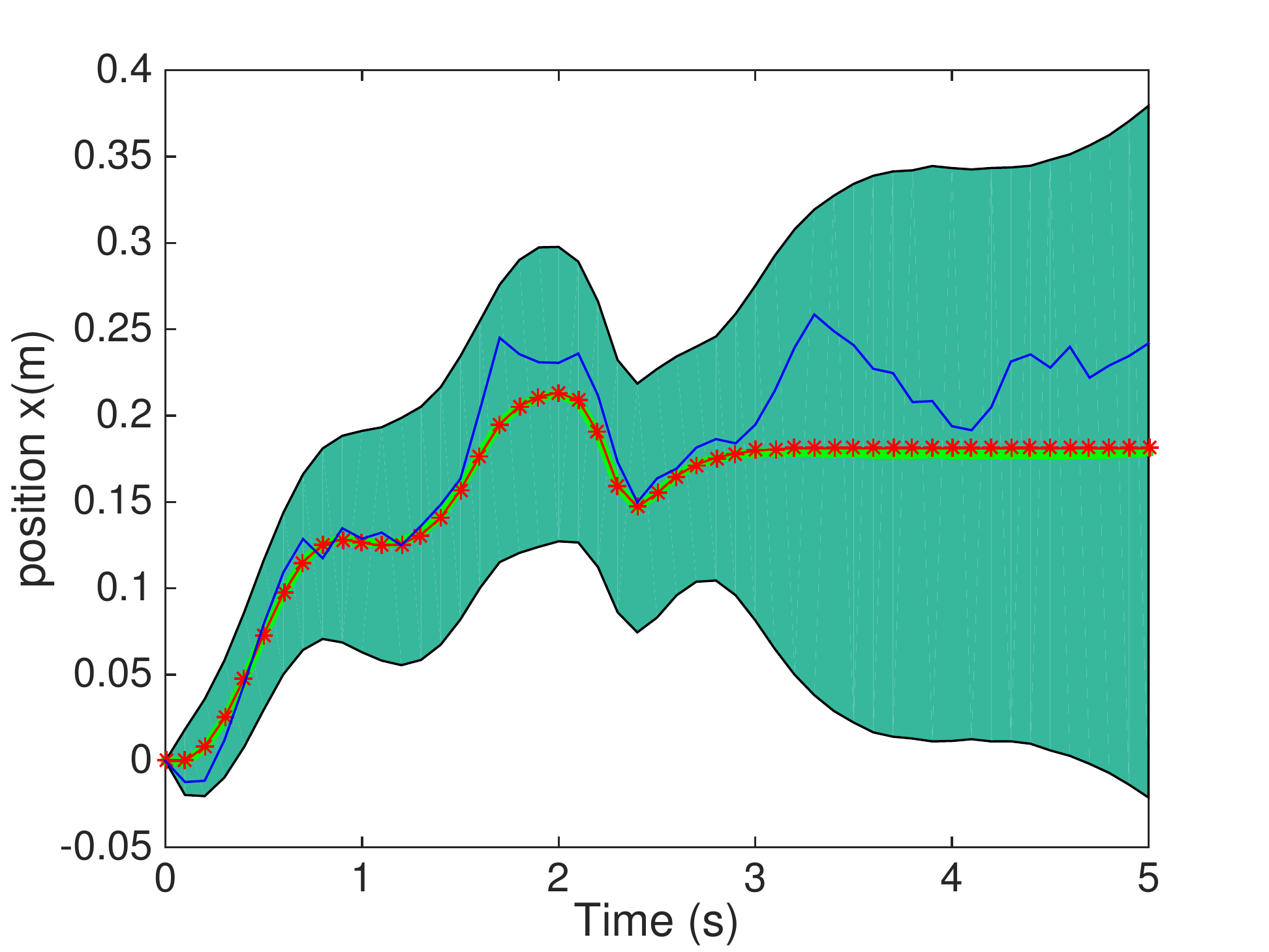}}
   \subfloat{\includegraphics[width= 0.32\textwidth]{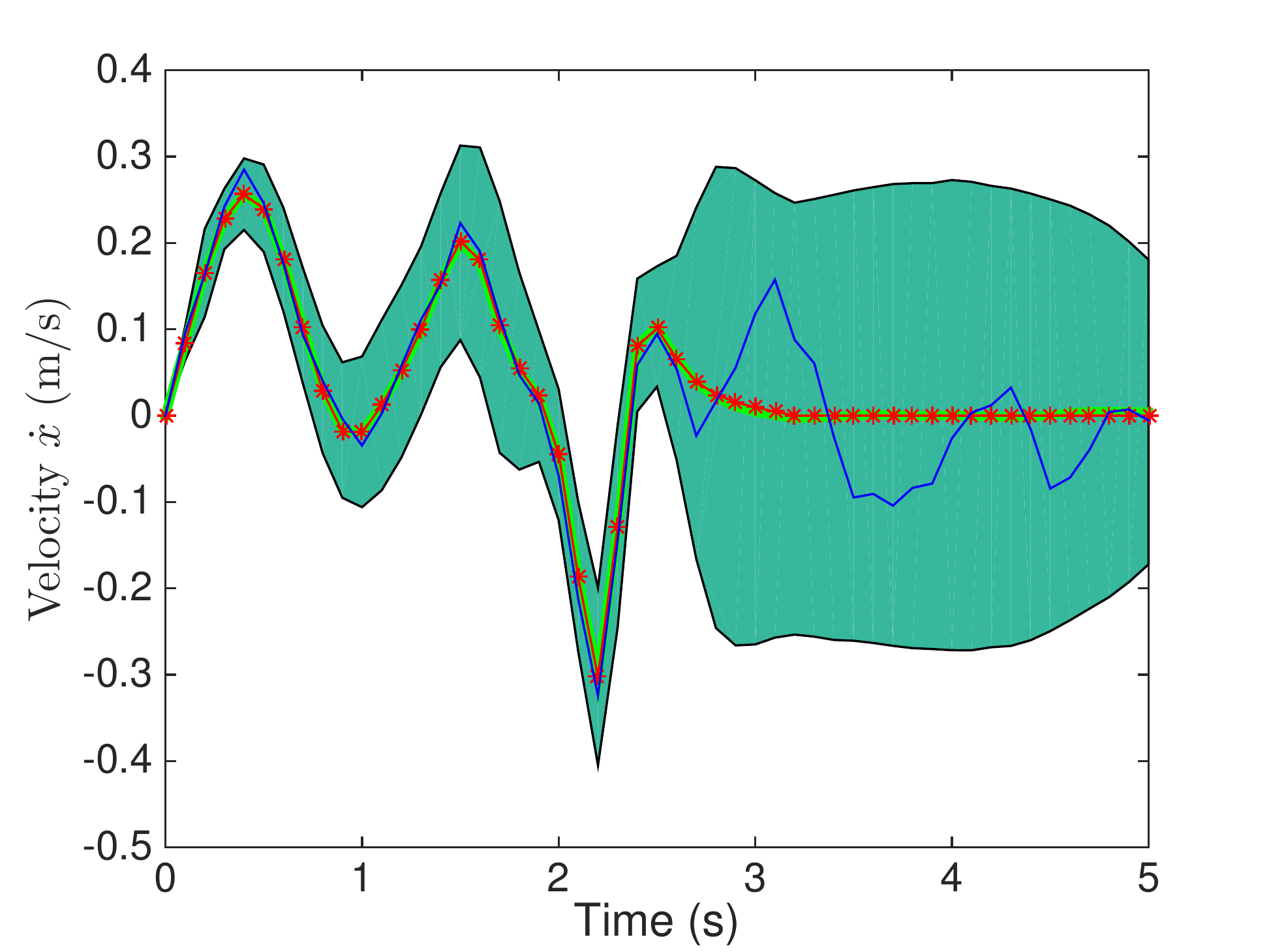}}
   \subfloat{\includegraphics[width= 0.32\textwidth]{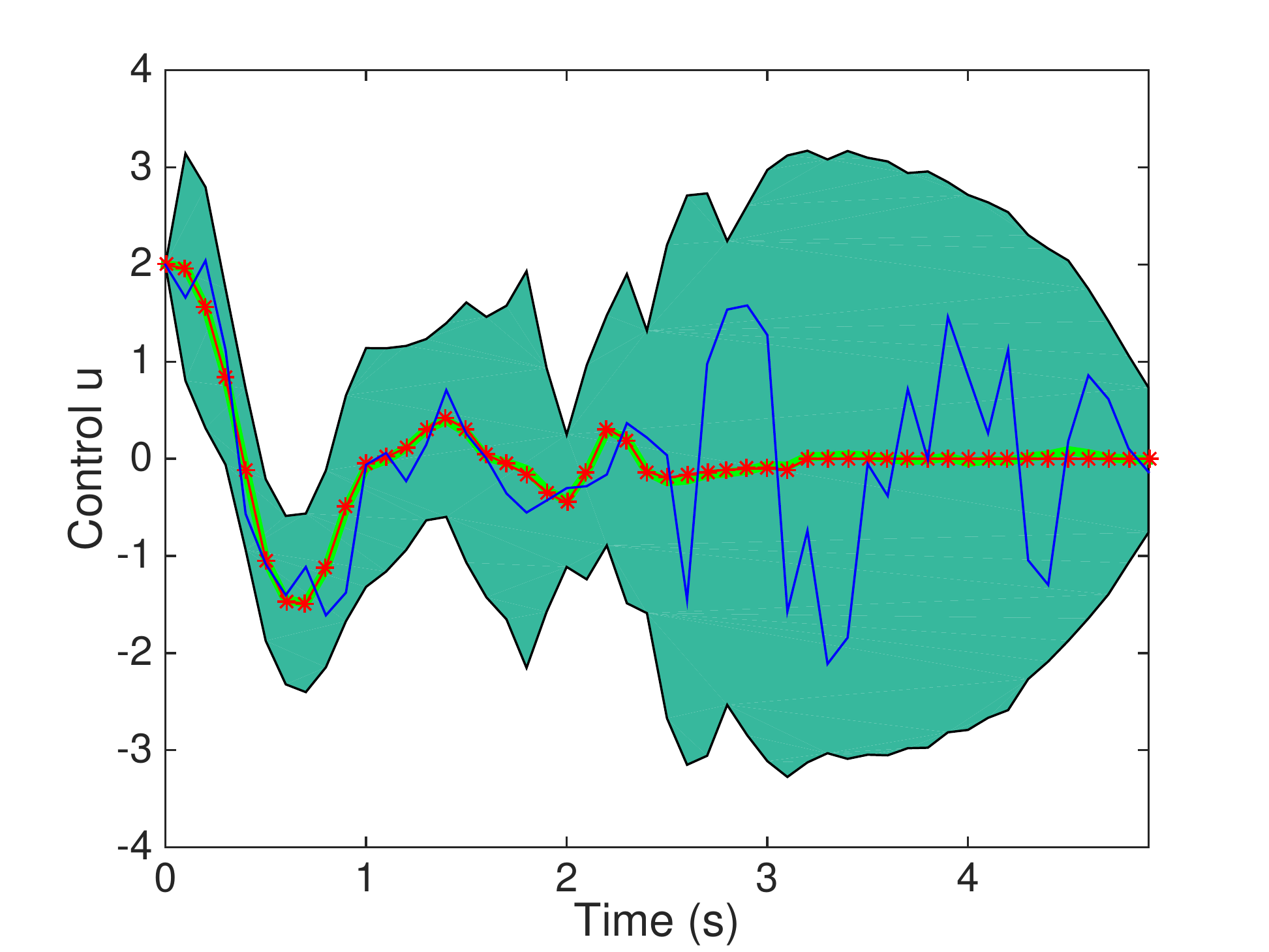}}
\caption{Performance of the closed-loop controller. The solid green lines represent the mean estimate, the  green shaded colors indicate the uncertainty with $2 \sigma$ bound, the red line with asterisk markers represent the open-loop nominal trajectory, and the blue line represents the actual states in one realization.\label{result}} 
\end{figure*}

The total simulation time is 5 seconds. The control objective is to find the control sequence $\{ u_k \}_{k = 0}^{N - 1}$ to make the pendulum swing up within 3.5 seconds and minimize the cost function $J_s$ in (\ref{cost_sto_orig}), and maintain the pendulum to the upright vertical position. The initial state $x_0 = [\pi; 0; 0; 0]$ is known, and the final state is $x_N = [0; 0; x_f; 0]$, where the cart position $x_f$ is not restricted. 
The open loop optimization problem is solved using Matlab nonlinear optimization solver $\mathtt{fminunc}$. The initial guess and optimal control are shown in Fig. \ref{result_open_opt}. The corresponding nominal trajectory $\{x_k\}_{k = 0}^{N}$ are plotted in Fig. \ref{result}. 

The implementation of Algorithm \ref{algo_ltv} to identify the linearized system is performed as follows.

The  size of the generalized Hankel matrix $H_k^{(p, q)}$ is $pn_y \times q n_u$, and as discussed before, the design parameters $p$ and $q$ should be chosen such that $\min \{pn_y, qn_u \} \geq n_x$, which for the current problem, $n_x = 4, n_u = 1,$ and $n_y = 4$. Design parameters $p, q$ are selected by trial and error. We start with some initial guess of $p, q$, compare the impulse responses of the original system and identified system, and check if the accuracy of the  identified system is acceptable.  Here, we choose $p = q = 5$.

We run $M$ parallel simulations to estimate the generalized Markov parameters $\{h_{k, j} \}_{j = 0}^{k}, k = 1, 2, \cdots, N -1$. We perturb the nominal control $\{\bar{u}_k \}_{k = 0}^{N - 1}$ with impulse, i.e., $\{\delta u_k^i \}_{k = 0}^{N - 1} = (0, 0, \cdots, 0.01, \cdots, 0)$ is the input perturbation sequence in the $i^{th}$ simulation, where only the $i^{th}$ element is nonzero. Therefore,  we choose design parameter $M = N$. In each simulation, we collect the outputs $\{\delta y_k^i \}_{k = 0}^{N-1}$  in (\ref{noise_free_output}) corresponding to the control input $\{\bar{u}_k + \delta u_k^i \}_{k  = 0}^{N - 1}$, and solve the least squares problem using (\ref{gmarkov}). 

\begin{figure}[!tb]
\centering
\includegraphics[width = 0.5 \textwidth]{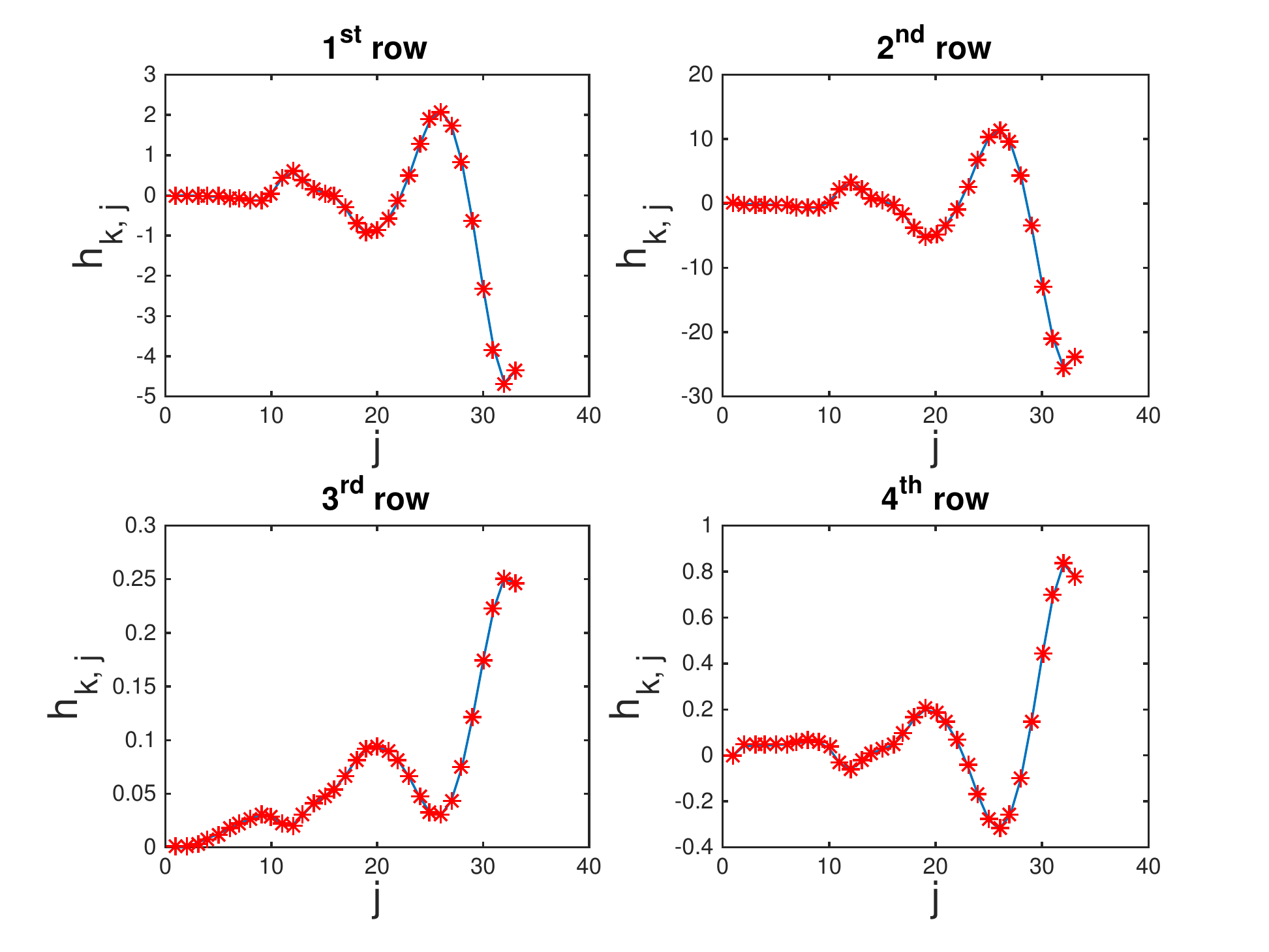}
\caption{Comparison of the generalized Markov parameters $\{h_{k, j} \}_{j = 0}^{k}$ as a function of time index $j$. The solid blue lines represent the identified Markov parameters, and the red asterisk markers represent the actual Markov parameters.\label{sim_markov}}\vspace{-5pt}
\end{figure}

We construct the generalized Hankel matrix, and solve the singular value decomposition problem. The rank of the Hankel matrix $n_r = 4$, and hence, the identified system $\hat{A}_k \in \Re^{4 \times 4}$. To test the accuracy of the identified system, we calculate the identified Markov parameters using $\hat{A}_k, \hat{B}_k, \hat{C}_k$, and compare with the actual generalized Markov parameters (calculated using impulse responses as mentioned above). Since the generalized Markov parameter $h_{k, j} \in \Re^{4 \times 1}$, all four elements are shown in Fig. \ref{sim_markov} for $k = 33$. 


With the identified linearized system, we design the closed-loop controller. We run 1000 individual simulations, where performance of the closed-loop controller is shown in Fig. \ref{result}. There are two observations: 1) the averaged state estimates over 1000 Monte-Carlo simulations runs (plotted in solid green lines) are close to the open-loop optimal trajectory (plotted in red), implying that the control objective to minimize the expected cost function could be achieved using the proposed approach; 2) in this problem, the closed-loop controller has better control in the angle and angular velocity, so that the corresponding $2 \sigma$ bound is tight. However, the uncertainties of position and velocity rise especially after the pendulum reaches the upright position. This is due to the fact that with the presence of process noise in all states, it is not possible to stabilize all the states simultaneously. 

\section{Conclusion}
In this paper, we have proposed a separation-based design of the stochastic optimal control problem for systems with unknown nonlinear dynamics and fully observed states in a separated open loop-closed loop fashion. First, we design a deterministic open-loop optimal trajectory.  Then we identify the nominal linearized system using time-varying ERA. The open-loop optimization and system identification are implemented offline, using the impulse responses of the system, and an LQG controller based on the ROM is implemented online.  The offline learning procedure is simple, and the online implementation is fast. We tested the proposed approach on the inverted pendulum problem, and showed the performance of the proposed approach. Future work will generalize the proposed approach to large-scale, partially observed systems.

\bibliographystyle{IEEEtran}
\bibliography{CDC_refs}

\end{document}